\documentclass[10pt,english]{article}
\usepackage{lmodern}

\usepackage[T1]{fontenc}
\usepackage[latin9]{inputenc}
\usepackage[letterpaper]{geometry}
\usepackage{babel}
\usepackage{array}
\usepackage{float}
\usepackage{multirow}
\usepackage{amsmath}
\usepackage{amssymb}
\usepackage{graphicx}
\usepackage{esint}
\usepackage{subscript}
\usepackage{float}
\usepackage{color}
\usepackage[bookmarks=false,linkcolor=red]{hyperref}
\hypersetup{colorlinks=true,breaklinks,urlcolor=[rgb]{0,0,1},citecolor=[rgb]{0,0.6,0}}
\usepackage{url}
\usepackage{multirow} 
\usepackage{array}
\usepackage[font=footnotesize,format=plain,labelfont=bf, margin=0.8cm]{caption}
\usepackage{url}
\usepackage{multirow}
\usepackage{array}
\usepackage{booktabs}
\usepackage{rotating}
\usepackage[numbers,sort]{natbib}
\usepackage{wrapfig}

\begin{document}

\title{Compressive Earth Observatory: An Insight from AIRS/AMSU Retrievals }

\author{Ardeshir M. Ebtehaj\textsuperscript{1}, Efi Foufoula-Georgiou\textsuperscript{2},
Gilad Lerman\textsuperscript{3} and Rafael L. Bras\textsuperscript{1}}

\maketitle
{\footnotesize{}\textsuperscript{1}School of Civil and Environmental
Engineering, Georgia Institute of Technology, Atlanta, Georgia, USA.}{\footnotesize \par}

{\footnotesize{}\textsuperscript{2}Department of Civil, Environmental, and Geo-Engineering,
University of Minnesota, Minneapolis, Minnesota, USA.}{\footnotesize \par}

{\footnotesize{}\textsuperscript{3}School of Mathematics, University
of Minnesota, Minneapolis, Minnesota, USA.}{\footnotesize \par}

\begin{abstract}

We demonstrate that the global fields of temperature, humidity and
geopotential heights admit a nearly sparse representation in the wavelet
domain, offering a viable path forward to explore new paradigms of
sparsity-promoting data assimilation and compressive recovery of land surface-atmospheric states from space.
We illustrate this idea using retrieval products of the Atmospheric Infrared Sounder (AIRS) and Advanced Microwave
Sounding Unit (AMSU) on board the Aqua satellite. The results reveal
that the sparsity of the fields of temperature is relatively pressure-independent
while atmospheric humidity and geopotential heights are typically sparser at lower
and higher pressure levels, respectively. We provide evidence that these land-atmospheric states can be accurately
estimated using a small set of measurements by taking advantage of their sparsity prior.

\end{abstract}

\section{Introduction}

Earth observations from space are an invaluable component for global
circulation models, reanalysis products, and regional/local predictive
models, especially in places where no ground observations are available for model
initialization and/or data assimilation. These spaceborne observations
are increasing at an unprecedented rate as new satellites
are launched and new missions planned in the next decade, such as
the Global Precipitation Measuring (GPM) mission whose core satellite
was launched in February 2014 \citep{Hou2013}, the Soil Moisture
Active Passive (SMAP) mission to be launched in late 2014 \citep{EntAL10}
and a series of other soon-to-follow missions (see, National Research
Council report on Earth Science and Applications from Space, A Midterm
Assessment of NASA's Implementation of the Decadal
Survey, 2012). In this paper we put forward the idea that efficient
acquisition, processing and assimilation of the land surface-atmosphere observations from space can
tremendously benefit by exploring their underlying spatial structure
via the recent advances in the theories of sparse approximation and Compressive Sensing \citep[and references therein]{Mal09, EldK12}.

A finite dimensional state vector representing a physical process is (nearly)
sparse in a certain domain if the amplitudes of a large number of
its representation coefficients are (nearly) zero in that domain.
Thus, a sparse discrete state vector can be well approximated using
only a few of its largest representation coefficients. Fourier and
wavelet transforms (a generalized finite differencing operator) are typically the gateway to reveal sparsity of
natural processes through representing them via a few elementary waveforms.

If the state variable of interest is sufficiently smooth, for example with derivatives of all orders,
then the Fourier decomposition is typically an effective sparsifying transform. However,
the Fourier decomposition yields dense representations with many non-zero
coefficients for piecewise smooth states that may contain transients
and localized events. Typically, regularity of these states, can be
sparsely captured in the amplitude of their wavelet coefficients \citep{Mal89}
and encoded as a priori knowledge to obtain improved solutions for
related inverse problems. Recent and fundamental developments in the
theory of sparse approximation and Compressive Sensing (CS) have offered
new directions enabling to obtain a highly accurate estimate of a sparse
state variable only from a small set of incoherent/random samples--much smaller than were classically required \citep{CanR06,CanRT06,CanRT06a,CanT06,Don06}.
In other words, rather than densely sampling a sparse state of interest,
the fundamental idea is to design a compressed sampling scheme
that allows us to acquire a small number of samples and then
accurately reconstruct the state via a sparsity-promoting convex optimization.

In land surface-atmosphere studies, these developments have inspired
recently proposed sparsity-promoting data assimilation methods to
address the analysis of sharp weather fronts \citep{FreNB12} and, in
a more general setting, to incorporate the sparsity of the underlying
state of interest in a transform domain for geophysical downscaling, data fusion
and assimilation problems \citep{EbtF12b,EbtZLF13,EfgEZa13}. In this
paper, we pursue two main goals using the AIRS/AMSU-A retrieval products.
First, we show that the global fields of temperature, humidity and
geo-potential heights are sparse in the wavelet domain, leveraging further developments
of the recently proposed sparsity-promoting variational data assimilation
approaches. Second, owing to the observed sparsity, we provide evidence
that these atmospheric and land-surface
states may be recovered from satellite measurements in a compressed form.

%% Figure 1
\begin{figure}[t]
\noindent \begin{centering}
\includegraphics[width=0.9\columnwidth]{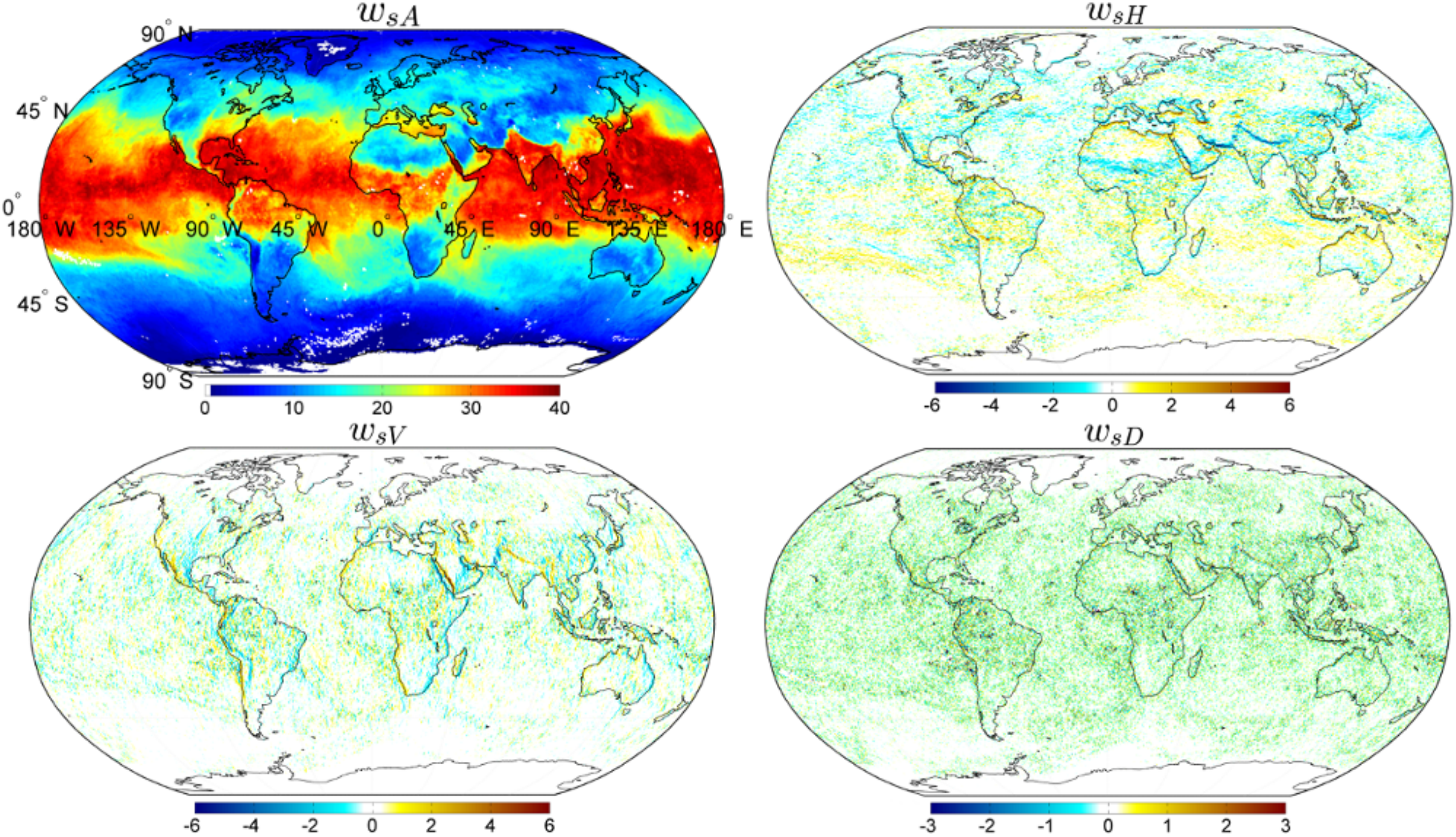}
\par\end{centering}

\caption{Haar wavelet coefficients at a single decomposition level for a three-day average of the AIRS level II standard
retrievals (AIRX2RET) of the surface water vapor mass mixing ratio
$w_{s}$ [g/kg dry air] from 09/01 to 09/03/2002--projected onto a 0.5-degree regular grid. Top panel from left-to-right:
low-pass $w_{sA}$ and high-pass horizontal wavelet coefficients $w_{sH}$.
Bottom panel from left-to-right: high-pass vertical $w_{sV}$ and
diagonal $w_{sD}$ wavelet coefficients, where the original moisture
field is $w_{s}=1/2\left(w_{sA}+w_{sH}+w_{sV}+w_{sD}\right).$ \label{fig:1}}
\end{figure}
%%--

\section{Evidence of Sparsity}

The distribution of energy and mass fluxes across the earth's
surface and its atmosphere is governed by complex physics that operate
at multiple scales of space and time. At the planetary scale, overturning
circulation of the tropical atmosphere gives rise to sharp moisture
gradients between the equatorial bands and dry subtropical ridges.
In synoptic-scale weather fronts, air masses differ by sharp transitions
in temperature and humidity, where the isolated highs and lows in
temperature-moisture fields are the main drivers of severe convective
activities and extreme weather. In high altitudes, near the tropopause,
the presence of jet streams may also give rise to discontinuities
and sharp transitions in moisture and temperature fields. In low altitudes,
near the earth surface, the spatial heterogeneity of the earth's
surface radiative forcing often manifests itself through sharp transitions
in surface temperature and moisture fields, especially over the land-ocean
interfaces, vicinity of snow-covered land surfaces and vegetation
regime changes. To accurately capture this multi-scale variability,
we traditionally need a dense uniform sampling pattern in space
and time that meets the Nyquist-Shannon sampling theorem. This theorem states that a continuous state variable can be exactly recovered from a set of uniformly spaced samples at the rate of at least twice the highest frequency content of the state.  It is then important to ask whether the underlying space-time structure of the land surface-atmospheric 
state variables of interest admits a sparse representation in an appropriate
domain, which can be subsequently exploited for accurate estimation, assimilation,
and speedy retrieval using much fewer samples than those required
by the Nyquist-Shannon rate. In this section, we provide evidence
that the spatial structure of the fields of temperature, humidity and geopotential
heights is markedly sparse in the wavelet domain, using the retrievals obtained from the instruments on board the Aqua satellite.

The Aqua satellite, launched in May 2002, carries the Atmospheric
Infrared Sounder (AIRS) and the Advanced Microwave Sounding Unit (AMSU-A)
among other instruments. This sensor package is one of the most advanced
integrated spaceborne hyperspectral instruments that scans the thermodynamic
structure of the earth's land-atmosphere with unprecedented accuracy
and space-time resolution \citep{PagAHO03}. The primary retrieval
products of the AIRS/AMSU-A include twice daily global fields of the
atmospheric temperature-humidity profiles among other cloud related
variables \citep{SusBB03,SusBIK11}. The AIRS/AMSU data have been
widely used for studying atmospheric thermodynamics \citep[among others]{Tia06,DuCF12},
improving operational weather forecasting via data assimilation \citep[among others]{LeMetal06,Wuetal12,Realeetal08,ReaEtal09,ReaLSR12}
and monitoring land surface hydrologic processes \citep[e.g.,][]{FerW10}.
In this paper, we confine our consideration to the version 6 of the
AIRS standard level-II retrieval products whose quality control indicators are zero (best) or one (good); see, AIRS/AMSU/HSB Version 6, Level 2 Product User Guide \citep{Ols13}. The daily level II data are stored in 240 granules, each includes 6 minutes of measurements registered onto 30 footprints across track by 45 lines along track with resolution of 45 km at nadir. Specifically, we study
the temperature-humidity profiles and geopotential heights, at standard pressure levels from
the earth surface up to 200 hPa. Throughout the paper, for illustration purposes, the data granules are projected onto a regular grid of 0.5-degree resolution.

%% Figure 2
\begin{figure}[t]
\noindent \begin{centering}
\includegraphics[width=0.8\columnwidth]{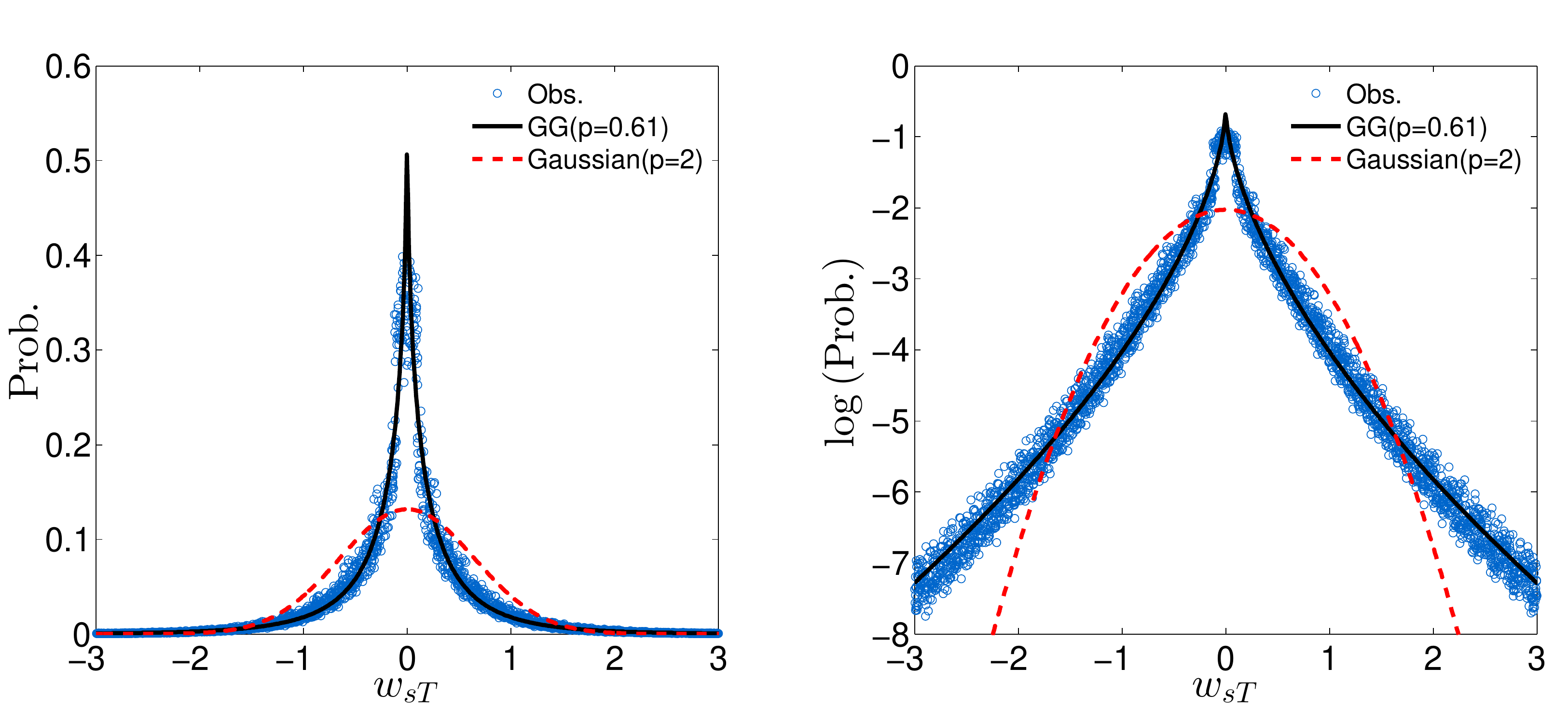}
\par\end{centering}

\caption{Left panel: the empirical probability masses (circles) of the wavelet
coefficients for the near surface water vapor mass mixing ratio (MMR) {[}g/kg dry air{]} obtained from 100 randomly sampled days in calendar years 2002-2014, each containing 240 orbital track granules of the standard AIRS level II retrievals (AIRX2RET). The solid and broken lines are the fitted Generalized
Gaussian (GG) distributions with parameter $p=0.61$ and $p=2$ (Gaussian
distribution), respectively. Here, $w_{sT}=\left[w_{sH},\, w_{sV},\, w_{sD}\right]$
is a vector, which stores all of the Haar high-pass wavelet coefficients
at a single decomposition level. The right panel demonstrates the
same on log-probability to contrast better the shape of the
probability masses versus the Gaussian density.\label{fig:2}}
\end{figure}
%%--

As explained, the wavelet decomposition provides a suitable transformation
that gives rise to a sparse representation for a state vector with local transitions 
and singularities. Here, we use the redundant discrete Stationary Wavelet Transform (SWT) \citep{NasS95}.
Owing to its redundancy and translation invariance, the SWT provides
a richer representation \citep{BodCK11} for sparse expression of
complex processes than the classic orthogonal wavelet decomposition
\citep{Mal89}. Figure \ref{fig:1} demonstrates the wavelet coefficients
of a global three-day average product (09/01 to 09/03/2002) of the level-II water vapor mass
mixing ratio (MMR) near the earth's surface. The three-day integration of data granules allows to
obtain an almost seamless estimate of the global MMR field with minimal averaging on the overlapping grids.

As is evident in Figure \ref{fig:1}, the high-pass wavelet coefficients
are markedly sparse as most of the coefficients are near zero. It
is seen that the horizontal wavelet coefficients capture the latitudinal moisture
variability while the vertical coefficients mostly encapsulate the
sharp zonal moisture transitions across land-ocean interfaces. The
large wavelet coefficients over ocean are mostly influenced by the
presence of sharp moisture gradients due to convective updrafts in
the Intertropical Convergence Zone (ITCZ) and downdrafts near the
subtropical ridges. Over land, large coefficients of surface moisture
are mainly concentrated near ocean-land interfaces and major mountainous
features covered with snow such as the Himalayas and Andes. In addition,
owing to the migration of the ITCZ over land, we see large horizontal
coefficients, for instance over the semi-arid Sahel, where the climate
and vegetation regime exhibits a sharp transition. Isolated moisture
highs can also be traced in the coefficients over the boundaries of
sufficiently large inland water bodies such as the Caspian Sea.

To study the presence of sparsity and characterize it as a priori knowledge, we randomly collected a
database containing 100 days of the AIRS/AMSU level II standard
retrieval products, from 2002 to 2014 (see Figure \ref{fig:S1} in the appendix). Figure \ref{fig:2} demonstrates the overlaid probability
masses (circles) for the high-pass wavelet coefficients of the near
surface water vapor MMR for all of the samples. We see that
the sparsity of the wavelet coefficients manifests itself as a large
probability mass around zero with extended tails much thicker than
the Gaussian density. In this figure the solid line is the fitted
Generalized Gaussian (GG) distribution with the following form:

\begin{equation}
\mathcal{P}_{X}\left(x\right)=\frac{p}{2\sigma\Gamma\left(1/p\right)}\exp\left(-\left|\frac{x}{\sigma}\right|^{p}\right),\label{eq:1}
\end{equation}
where the Gamma function is $\Gamma\left(z\right)=\int_{0}^{\infty}e^{-t}\, t^{z-1}\, dt$
for $z>0$ and the non-negative parameters $p$ and $\sigma$ determine
the shape and width of the density, respectively \citep[see,][]{Nad05}.
Evidently, this family of distributions is log-concave for $p\geq1$
and contains the well-known Gaussian ($p=2$) and Laplace ($p=1$)
densities as special cases. Here, we consider the shape parameter
$p$ as a measure that characterizes the degree of sparseness, noting
that the density tends to the Dirac delta function with maximum nominal
sparsity for $p\rightarrow0$. It is observed that not only the moisture
fields but also the temperature and pressure fields are sparse
in the wavelet domain and can be well explained by the GG density
with a much ticker tail than the Gaussian distribution (Figure \ref{fig:S2}-to-\ref{fig:S4}
in the appendix). To concisely estimate the shape parameter
and thus the sparsity of the wavelet coefficients, we note that the
following ratio of the first and second order moments

\begin{equation}
\mathcal{M}\left(p\right)=\frac{\mathbf{m}_{\mathbf{2}}}{\mathbf{m}_{\mathbf{1}}^{\,\,\,2}}=\frac{\Gamma\left(3/p\right)\Gamma\left(1/p\right)}{\left|\Gamma\left(2/p\right)\right|^{2}},\label{eq:2}
\end{equation}
is only a function of the shape parameter in the GG density, where

\begin{equation}
\mathbf{m}_{n}=\int_{-\infty}^{\infty}\left|x\right|^{n}\mathcal{P}_{X}(x)\, dx.\label{eq:3}
\end{equation}

%% Figure 3
\begin{figure}[h]
\noindent \begin{centering}
\includegraphics[width=0.92\columnwidth]{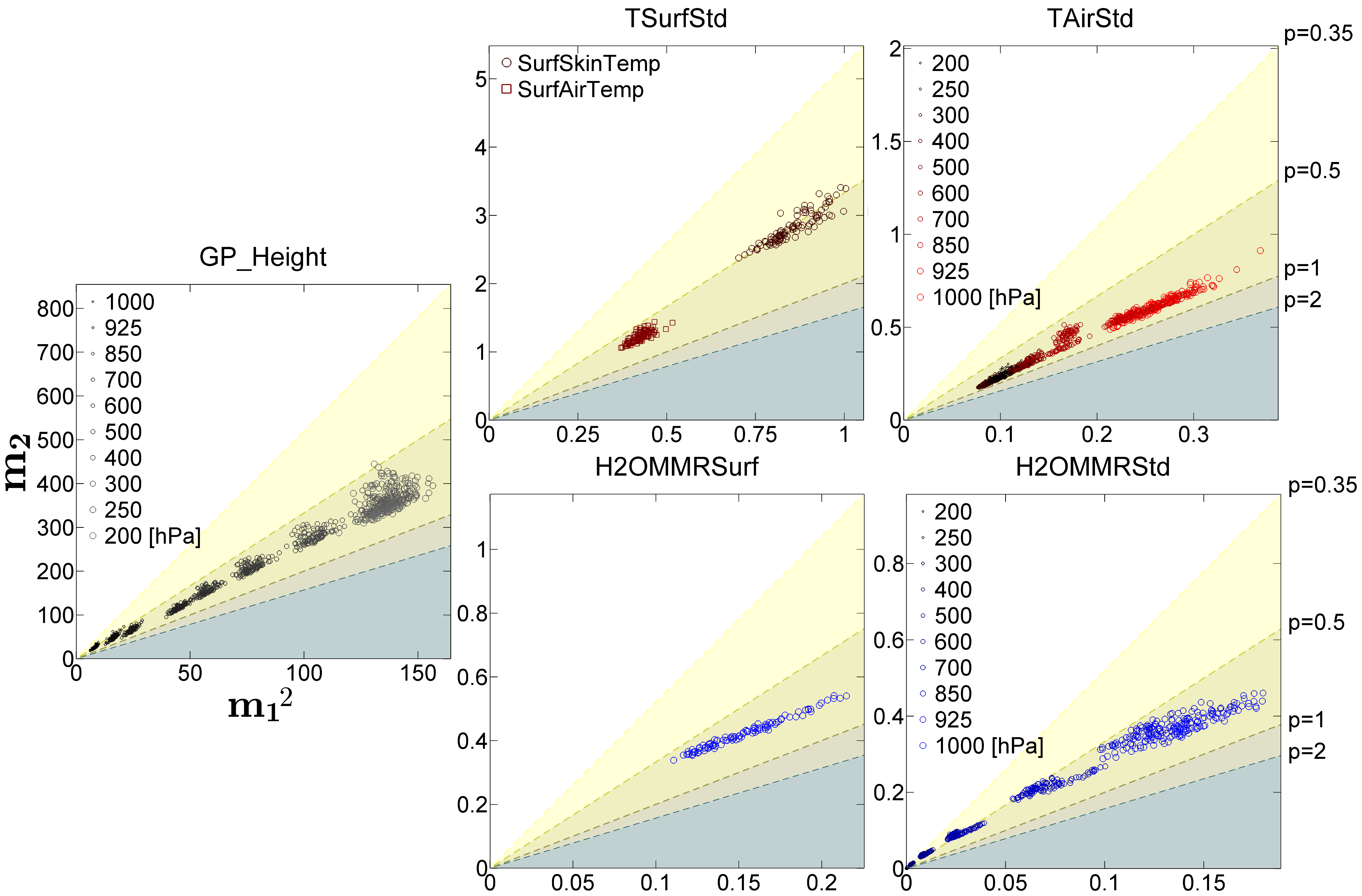}
\par\end{centering}

\caption{Demonstration of sparsity of the wavelet coefficients of the global
fields of geopotential heights {[}meters{]}, temperature {[}Kelvin{]}
and water vapor mass mixing ratio {[}g/Kg dry air{]} through a statistical
moment diagram. The moments are computed from 100 randomly sampled days of AIRS
level II standard retrievals, from the earth surface up to 200
hPa. Top-to-bottom panels from left-to-right are: the first (x-axis)
and second (y-axis) order moment-pairs of the wavelet coefficients
for the fields of geopotential heights (GP\_Height); surface (TSurfStd)
and profiles (TAirStd) of the atmospheric temperatures; and surface
(H20MMRSurf) and profiles (H2OMMRStd) of the water vapor mass mixing
ratios. The lines with slope $\mathcal{M}\left(p\right)$ (see equation
\ref{eq:2}) partition the moment space into sparser ($p\leq1$) and
denser ($p>1$) representations in the wavelet domain.
All of the studied fields exhibit remarkable sparsity in the
wavelet domain as in all cases $p\leq1$.\label{fig:3}}
\end{figure}
%%--

Figure \ref{fig:3} demonstrates the first versus the second-order
moments of the high-pass wavelet coefficients of the geopotential
heights {[}meters{]}, temperature {[}Kelvin{]} and moisture MMR {[}g/kg
dry air{]} fields, for all of the retrievals over the 100 sampled days. In this figure, the shaded areas stratify
the probability continuum of the GG density for sparser ($p\leq1$) and denser ($p>1$) representation of the wavelet coefficients in terms
of the moment ratio in equation (\ref{eq:2}). We can see that for
all states of interest, the sparsity is measured by $p\leq1$ throughout
the studied atmospheric depth (see, Table \ref{tab:S1} in the appendix).
For the geopotential heights, the magnitude of the moment pairs and
their spread grow almost linearly from high to low pressure levels, while higher pressure fields are slightly sparser than the low-pressure ones (Figure \ref{fig:3}, middle-row
panel). The sparseness of the temperature fields is relatively pressure-independent
(Figure \ref{fig:3}, top-row panels). As the air temperature contains more frequent sharp transitions near the earth
surface, we consistently see larger moments of the wavelet coefficients at higher-pressure levels.
As is evident, surface skin and air temperature fields are among the sparsest fields with
$p\cong0.5$. It is seen that the surface skin temperatures exhibit stronger
sparsity with slightly smaller shape parameter $p$, compared to the near surface air temperatures.
This observation reflects the fact that the fields of surface skin
temperatures contain sharper transitions and thus larger wavelet coefficients,
compared to the surface air temperatures, partly because of the mixing
and diffusive effects of the planetary boundary layer. The moisture
fields are also sparse but their sparsity exhibits a notable pressure-dependence
(Figure \ref{fig:3}, bottom-row panels). Specifically, it
can be seen that moisture fields at higher-pressure levels are less
sparse with larger $p$ values than in the upper atmosphere.
In other words, compared to the moist lower atmosphere, it seems that
the spatial distribution of moisture in the dry upper atmosphere is
relatively invariant in space, giving rise to a large number of near
zero fluctuations captured by the wavelet coefficients. However, the upper atmosphere moisture fields perhaps contain localized and sharp transitions, partly due to intermittent deep convections and movements of the jet streams, giving rise to a heavier tail and sparser distribution of the wavelet coefficients.

\section{Compressive Sensing of Land Atmospheric States}

Owing to the sub-hourly and sub-kilometer evolution of the mesoscale
land surface-atmospheric dynamics, spaceborne remote sensing at the
Nyquist-Shannon rate seems costly and often infeasible for now, at least
from a hardware perspective. In this section, we provide a brief introduction
to Compressive Sensing (CS) and then present experimental evidence that
owing to the observed sparsity, the global fields of temperate and
humidity can be recovered with sufficient accuracy using only a few incoherent/random samples.

As previously mentioned, CS is an emerging field
in statistical estimation theory that allows to reconstruct sparse
state vectors only from a few randomized measurements. Specifically, for a perfect
reconstruction of a continuous state of interest from its discrete
samples, the classic Nyquist-Shannon sampling theorem demands a uniform
sampling rate of at least twice the highest frequency content of the
state variable of interest. However, central results of CS suggest
that we can recover a sparse state from a much smaller set of measurements
than those required by the Nyquist-Shannon criterion. In particular,
let us assume that the state $\mathbf{x}\in\mathfrak{R}^{m}$ is represented by an $m$-element vector which has $k$ non-zero
elements, either in the ambient or a suitable transform domain (e.g.,
wavelet). Furthermore, let us consider that a set of under-sampled
measurements $\mathbf{y}\in\mathfrak{R}^{n}$, that is $n\ll m$, are related to the true
state $\mathbf{x}$, through the
following linear model

\begin{equation}
\mathbf{y=}\mathbf{Hx}+\mathbf{v},\label{eq:4}
\end{equation}
where $\mathbf{v}\in\mathfrak{R}^{n}$ represents an error with an
$n$-by-$n$ covariance matrix $\mathbf{R}\in\mathfrak{R}^{n\times n}$,
and $\mathbf{H}\in\mathfrak{R}^{n\times m}$ denotes a specifically
designed ``sensing matrix'' that linearly samples the state variable
of interest. Notice that rather than sampling the state uniformly
at specific points in time and space, CS "samples" it as inner products
between the state of interest and rows of the linear sensing matrix $\mathbf{H}$. The CS theory proves \citep[see,][]{CandTao05} that we can recover, with high-degree of accuracy, the state of interest from a few randomly
chosen linear measurements $\mathbf{y}$ via the $\ell_{1}$-norm regularization
of a classic least-squares estimator as follows:

\begin{equation}
\underset{\mathbf{x}}{\text{minimize}}\;\left\{ \left\Vert \mathbf{y}-\mathbf{H}\mathbf{x}\right\Vert _{\mathbf{R}}^{2}+\lambda\left\Vert \mathbf{W}\mathbf{x}\right\Vert _{1}\right\} ,
\label{eq:5}
\end{equation}
where, the $\ell_{1}$-norm is defined by $\left\Vert \mathbf{x}\right\Vert _{1}=\Sigma_{i=1}^{m}\left|x_{i}\right|$,
the quadratic norm is $\left\Vert \mathbf{x}\right\Vert _{\mathbf{R}}=\mathbf{x}^{T}\mathbf{R}^{-1}\mathbf{x}$,
$\mathbf{W}$ represents a suitable sparsifying transformation (e.g.,
wavelet) and $\lambda>0$ is a regularization parameter. Minimization
of the $\ell_{1}$-norm promotes sparsity while the quadratic penalty
ensures fidelity of the solution to the observations. Note that from
the Bayesian statistical point of view, the $\ell_{1}$-norm regularization
is equivalent to assuming that the underlying sparsity can be a priori
explained by the GG density with $p=1$.

The design of the sensing matrix $\mathbf{H}$ plays a critical role in the success of a CS reconstruction. Practically, this matrix has to sample the underlying state $\mathbf{x}$ incoherently to assure a quasi-uniform spreading of the ``uncaptured energy'' across the entire field, for avoiding aliasing and blurring artifacts in the reconstruction process \citep{Lus08}.
Theoretically, the CS recovery in problem (\ref{eq:5}) succeeds with high probability when the matrix $\mathbf{H} \mathbf{W}^T$ (which is the multiplication of $\mathbf{H}$ with the transpose of $\mathbf{W}$) behaves like an orthogonal transformation when operating on a sparse state. Such a behavior has been precisely characterized by a few mathematical notions, including the Restricted Isometric Property (RIP) by \citet{CandTao05}, and the Spark and Mutual-Coherence by \citet{DonE03}. In our analysis, the sensing matrix $\mathbf{H}$ is obtained by random sampling of a few rows of the identity matrix, which is equivalent to obtaining the observations from a small number of pixels in the satellite field of view. While, CS problem (\ref{eq:5}) often succeeds with this incoherent sensing matrix, $\mathbf{H} \mathbf{W}^T$ cannot be guaranteed to have the RIP and the theoretical study of similar sampling schemes is a topic of an ongoing research \citep{RauhutW12, Karhmer_Ward14}

%% Figure 4
\begin{figure}
\begin{centering}
\includegraphics[width=0.9\columnwidth]{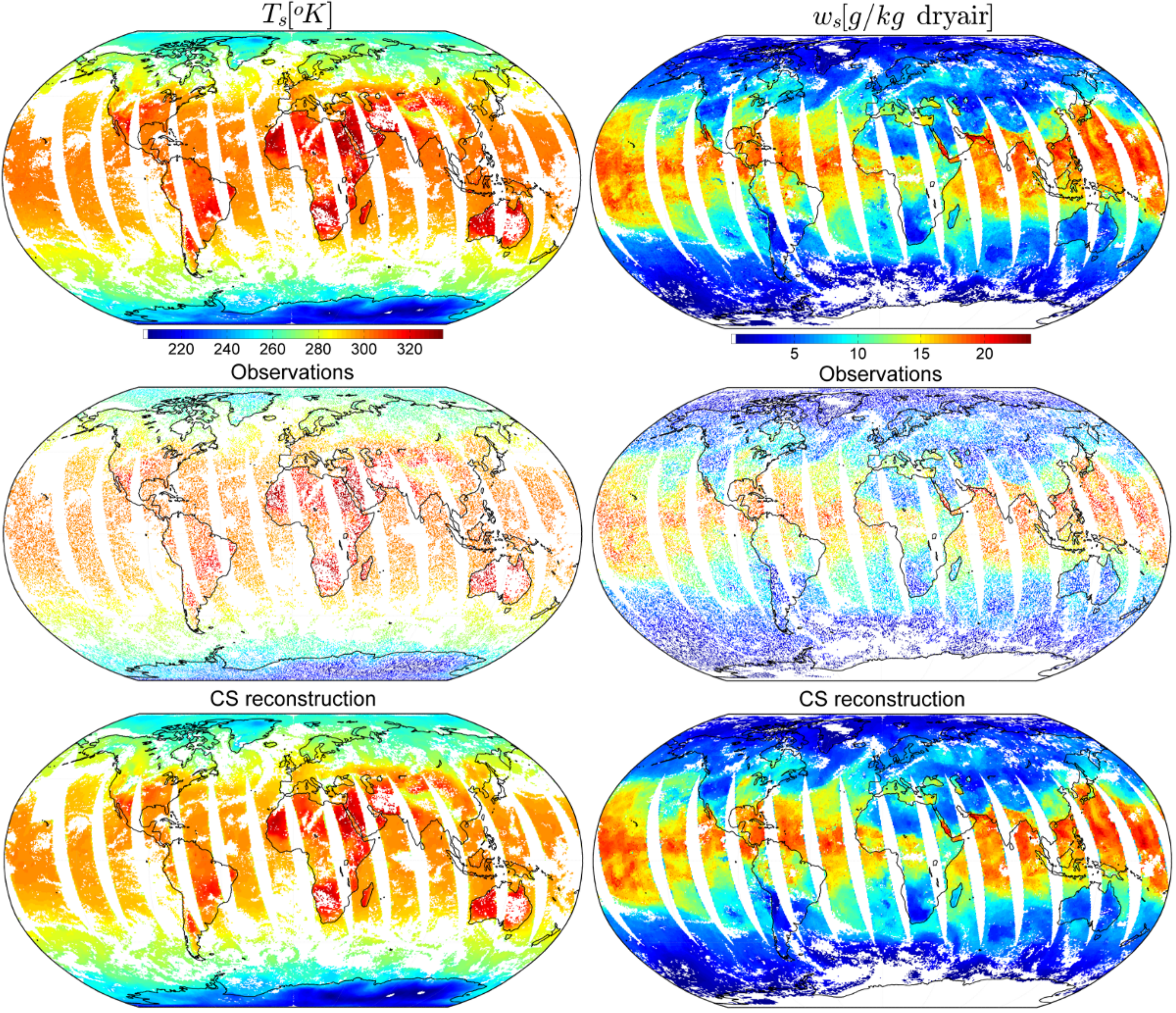}
\par\end{centering}

\caption{CS reconstruction of the surface skin temperature {[}Kelvin{]} (left column, 01/01/2002) and
surface water vapor mass mixing ratio {[}g/kg dry air{]} (right column, 09/09/2002) for all daily ascending orbital tracks.
The originally retrieved fields (top row), the randomly
under-sampled fields (middle row), and the CS reconstructed fields (bottom row).
The under-sampled fields only contain 45\% of randomly chosen pixels, which are also corrupted with a
white Gaussian noise.\label{fig:4}}
\end{figure}

The top panels in Figure \ref{fig:4} demonstrate ascending granules of the
AIRS/AMSU level II standard retrievals of the surface skin temperature
on 01/01/2002 and near surface water vapor MMR on 09/09/2002, respectively. In these figures, the data granules are projected onto a regular grid with 0.5-degree resolution using the nearest-neighbor interpolation. To implement the CS reconstructions, all the pixels of the original fields were first scaled into the range between 0
and 1. Then, we randomly sampled only 45\% of the retrieved pixels
and added a small zero-mean white Gaussian noise with standard deviation
$10^{-3}$ to those samples (Figure \ref{fig:4} middle panels).
The results of the CS reconstruction are demonstrated in the bottom
panels of Figure \ref{fig:4}. In our experiments, we used the Fast
Iterative Shrinkage-Thresholding Algorithm by \citet{BecT09a}. The
regularization parameter in problem (\ref{eq:5}) needs to be determined
empirically. We chose $\lambda=0.02$, which we found to work well
for our case studies (see, \citet{KimKLB07} for feasible ranges of
$\lambda$). Note that, owing to the chosen sensing matrix, the initial energy (sum of squares) of the error in these experiments is approximately
55\% of the total energy of the original fields. After CS reconstruction
the energy of error is less than 0.03\% and 2\% of the total energy in the examined
temperature and moisture fields, respectively. Although CS recovers
remarkably well the large-scale features of the moisture-temperature
fields, a close inspection shows that some isolated moisture highs
have not been well captured in the reconstruction process (e.g., over the Red Sea), especially near the swath edges.
We empirically observed that under the same sampling scheme, typically, the surface temperature
fields can be recovered with less reconstruction error than that
of the moisture fields, which may be partly due to the stronger sparsity
prior (smaller $p$ values) in the temperature fields (Table \ref{tab:S1} in
the appendix). We also found that for a randomized sample
size of more than 50-to-60\% of the total pixels,
the reconstruction is almost perfect and CS recovers well even the small-scale
features of the moisture fields. Typically, the reconstruction quality
severely degrades, and it may diverge, for sample sizes of less than
20\% of the total pixels.

\section{Concluding Remarks}

We have provided compelling evidence that the spatial structure of
the fields of temperature-moisture and geopotential heights is sparse
in the wavelet domain. This result is key for supporting further
developments of the recently suggested sparsity-promoting data assimilation
methods. Furthermore, exploiting the sparsity prior, we provided promising
results on the use of Compressive Sensing (CS) theory for efficient reconstruction
of these primary state variables from a few linear random measurements/retrievals.
While progress has been made recently in developing sparse
digital image acquisition in visible bands \citep{DuaDB08}, development
of sparse-remote-sensing instruments for earth observations from space,
in microwave and infrared wavelengths, remains an important challenge
in the coming years. However, our results suggest that, even under
the current sensing protocols, only a few randomly chosen pixel-samples
of the primary land-atmospheric states can be advantageously exploited for a speedy reconstruction
of the entire sensor's field of view with a notable degree of accuracy. Similar to the prominent applications of CS in rapid Magnetic Resonance Imaging \citep{Lus08},
further research is needed to explore practical randomized scanning strategies,
compatible with the currently used scanning geometries (e.g., conical, across track),
to advance applications of CS for more efficient and faster retrievals of geophysical
states from space. Clearly, the implications of such a capability cannot be
overstated for real-time tracking and data assimilation of extreme land-atmospheric
phenomena in global early warning systems.

%----------------------------------------------------------------------------------------
%	ACKNOWLEDGEMENTS
%----------------------------------------------------------------------------------------

\section{Acknowledgment}

The authors acknowledge support provided by two NASA Global Precipitation
Measurement grants (NNX13AG33G and NNX13AH35G) and an NSF grant (DMS-09-56072).
The support by the K. Harrison Brown Family Chair and the Joseph T.
and Rose S. Ling chair is also gratefully acknowledged. The AIRS/AMSU-A (version 6) data were obtained from the NASA's Goddard Earth Sciences Data and Information Services Center (GES DISC), which are freely accessible at \url{http://disc.sci.gsfc.nasa.gov/AIRS/data-holdings}. The authors also would like to thank Dr.~Jarvis Haupt at the University of Minnesota for his offered insights and helpful discussions.

\section{Appendix}
%% APPENDIX
\appendix
\renewcommand\thefigure{\thesection.\arabic{figure}} 
\renewcommand\thetable{\thesection.\arabic{table}} 
\setcounter{figure}{0} 
\setcounter{table}{0} 
\numberwithin{equation}{section}

This appendix contains some figures and descriptions, which are important
for clarity and further support of the main arguments of the paper.
Specifically, using the standard AIRS/AMSU level II retrievals, this
document provides additional evidence supporting the idea that the
global fields of temperature-humidity and geopotential heights are
sparse in the wavelet domain. Furthermore, we provide complementary
descriptions regarding to the design and interpretation of the presented
Compressive Sensing (CS) reconstruction experiments.

\section{Evidence of Sparsity\label{sec:S1}}

\begin{figure}
\noindent \begin{centering}
\includegraphics[width=0.5\paperwidth]{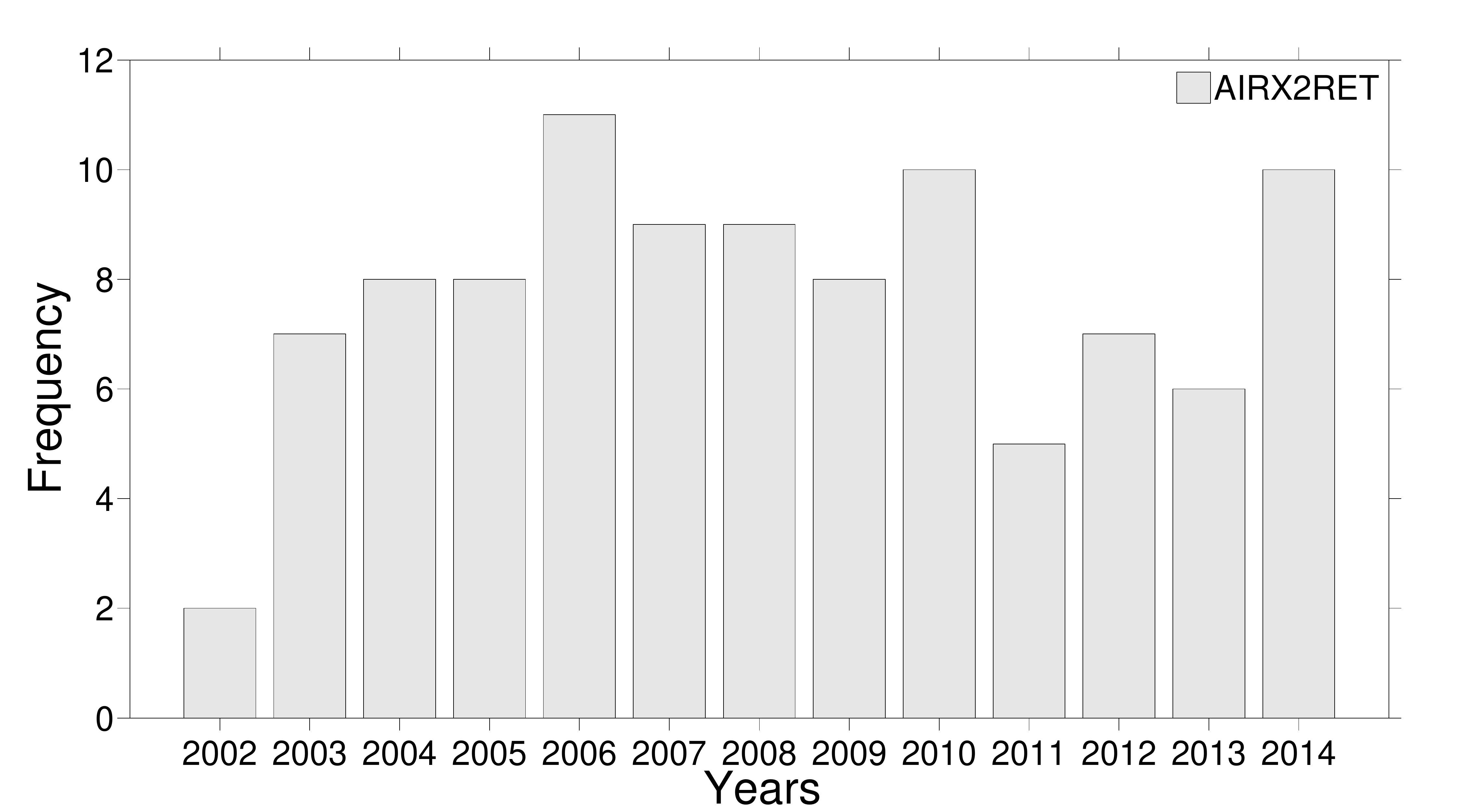}
\par\end{centering}

\protect\caption{Annual frequency histogram for 100 sampled days of the AIRS level II
standard retrievals in calendar years 2002 to 2014. The sampled data are used to infer sparsity of the fields of geopotental height, temperature and moisture at different pressure levels. \label{fig:S1}}
\end{figure}

\begin{figure}
\begin{centering}
\includegraphics[width=0.35\paperwidth]{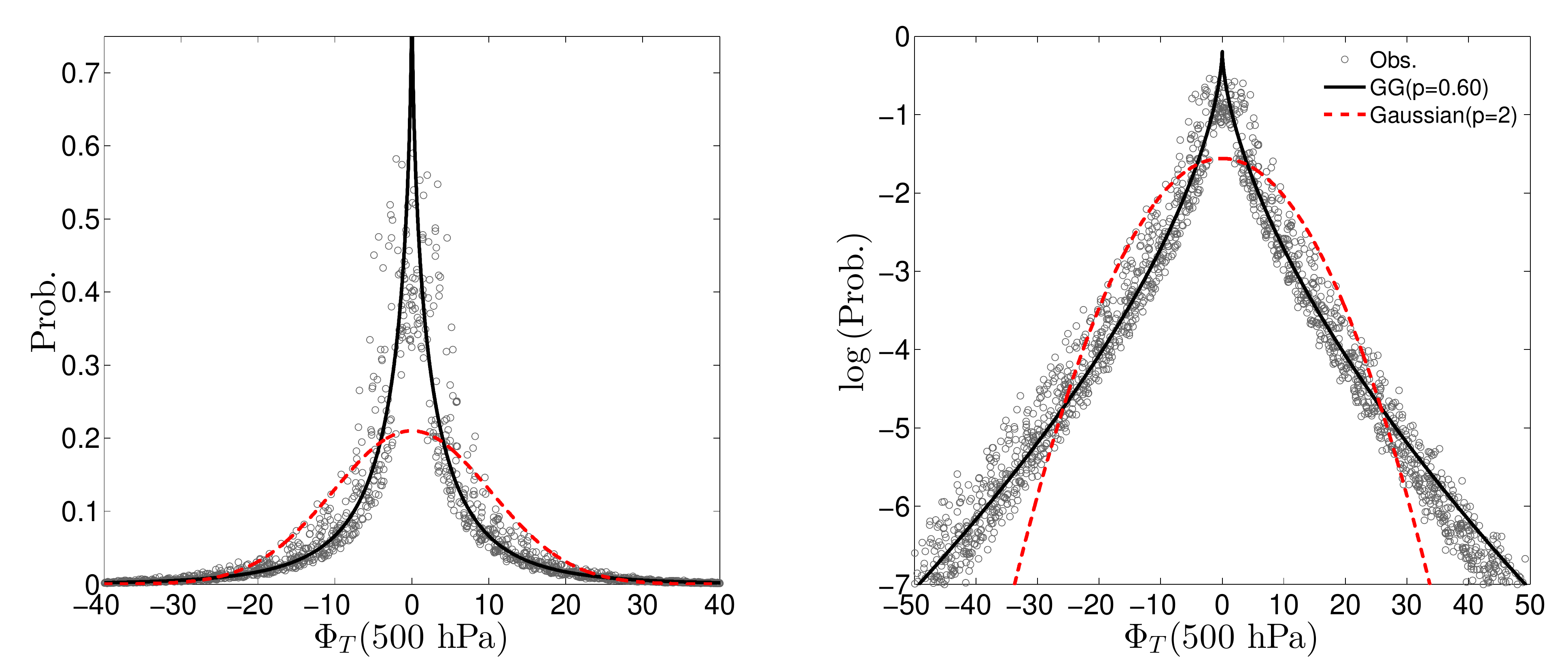} \includegraphics[width=0.35\paperwidth]{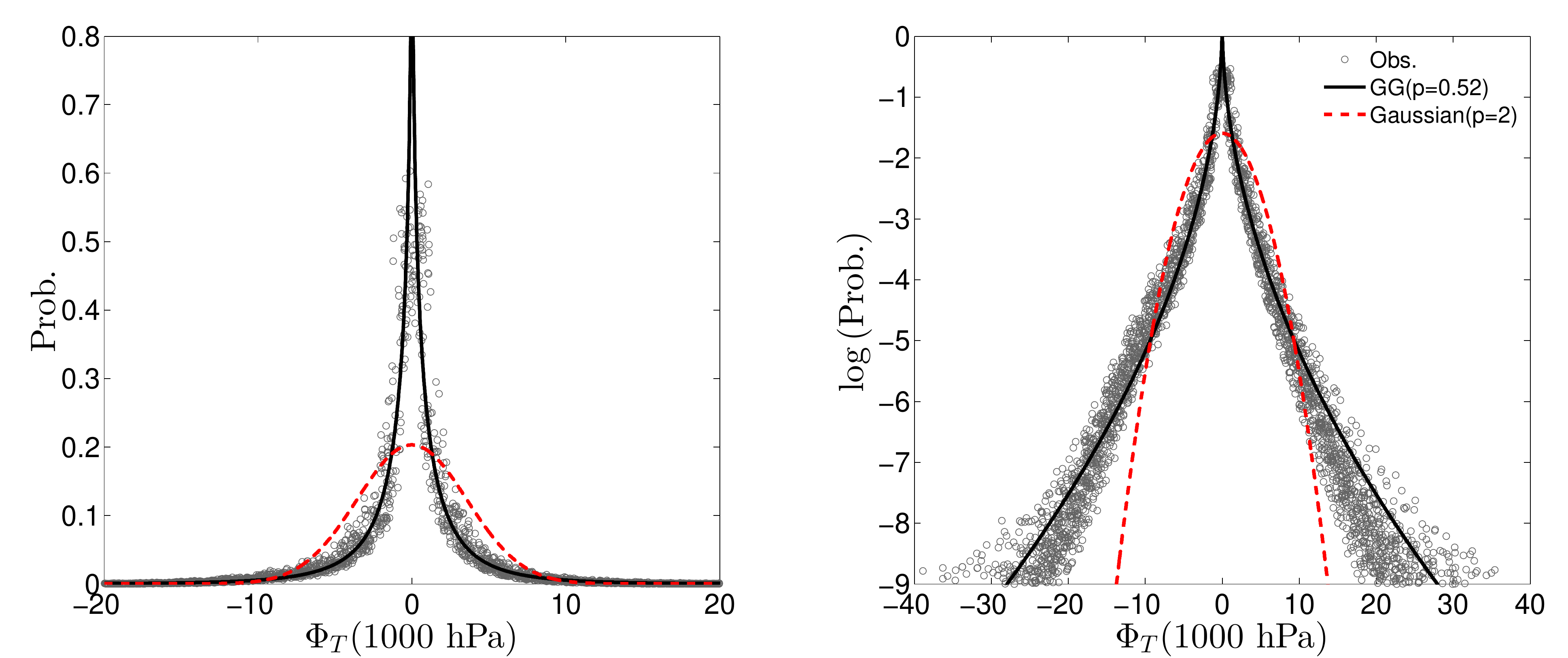}
\par\end{centering}

\protect\caption{Sparsity in the fields of geopotential height. Left-to-right panels:
the overlaid empirical probability masses (circles) of the wavelet
coefficients for the geopotential heights {[}m{]} at
500 hPa (left panel) and 1000 hPa (right panel).
The results are obtained from the 100 sampled daily retrievals of the AIRS level II products (see, Figure \ref{fig:S1}). Here, $\Phi_{T}=\left[\Phi_{H},\,\Phi_{V},\,\Phi_{D}\right]$
stacks all of the horizontal, vertical and diagonal high-pass wavelet
coefficients at a single decomposition level. The solid (black) and broken (red)
lines are the fitted GG and Gaussian distributions, respectively.\label{fig:S2}}
\end{figure}

\begin{figure}
\begin{centering}
\includegraphics[width=0.35\paperwidth]{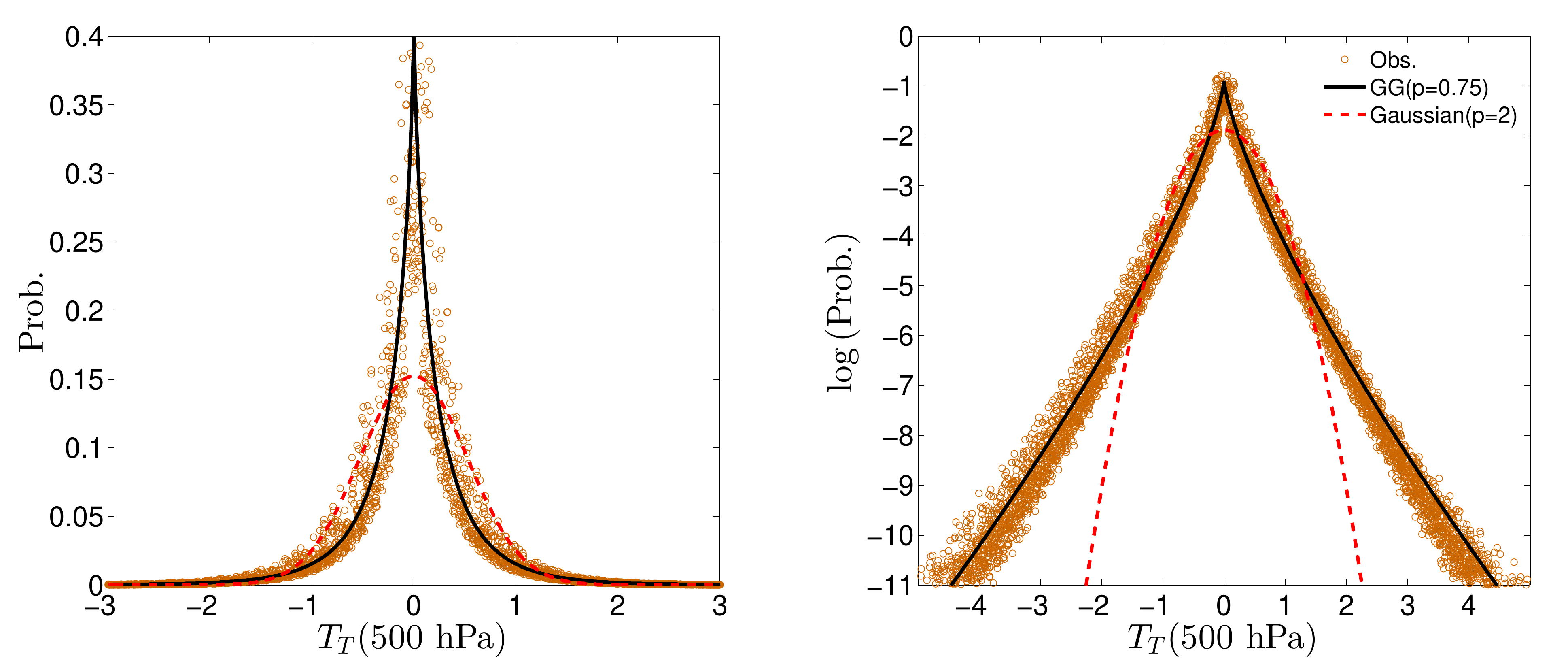} \includegraphics[width=0.35\paperwidth]{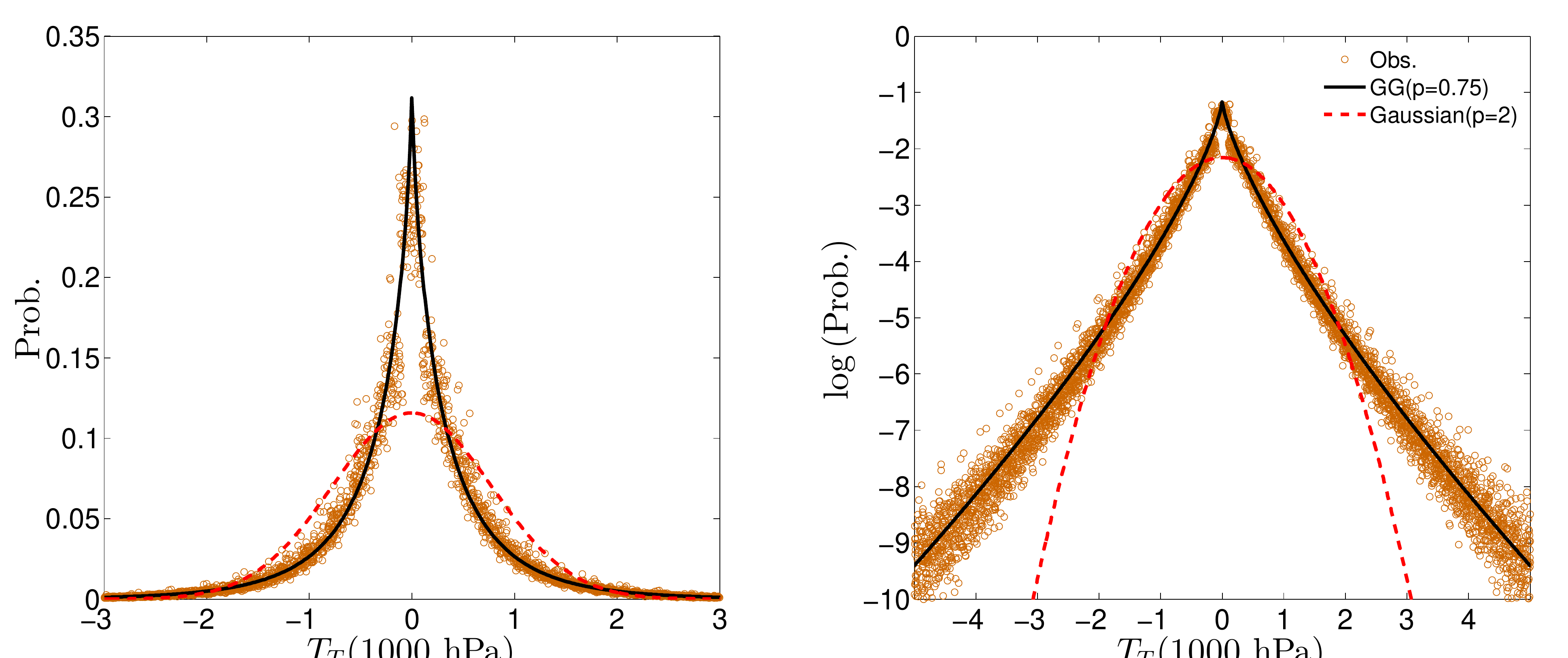}
\par\end{centering}

\protect\caption{Sparsity in the temperature fields. The overlaid empirical
probability masses (circles) of the wavelet coefficients for the air
temperature {[}Kelvin{]} fields at pressure levels 500 hPa (left panel)
and 1000 hPa (right panel). Please see caption of Figure \ref{fig:S2}
for relevant explanations. \label{fig:S3} }

\end{figure}

\begin{figure}
\begin{centering}
\includegraphics[width=0.35\paperwidth]{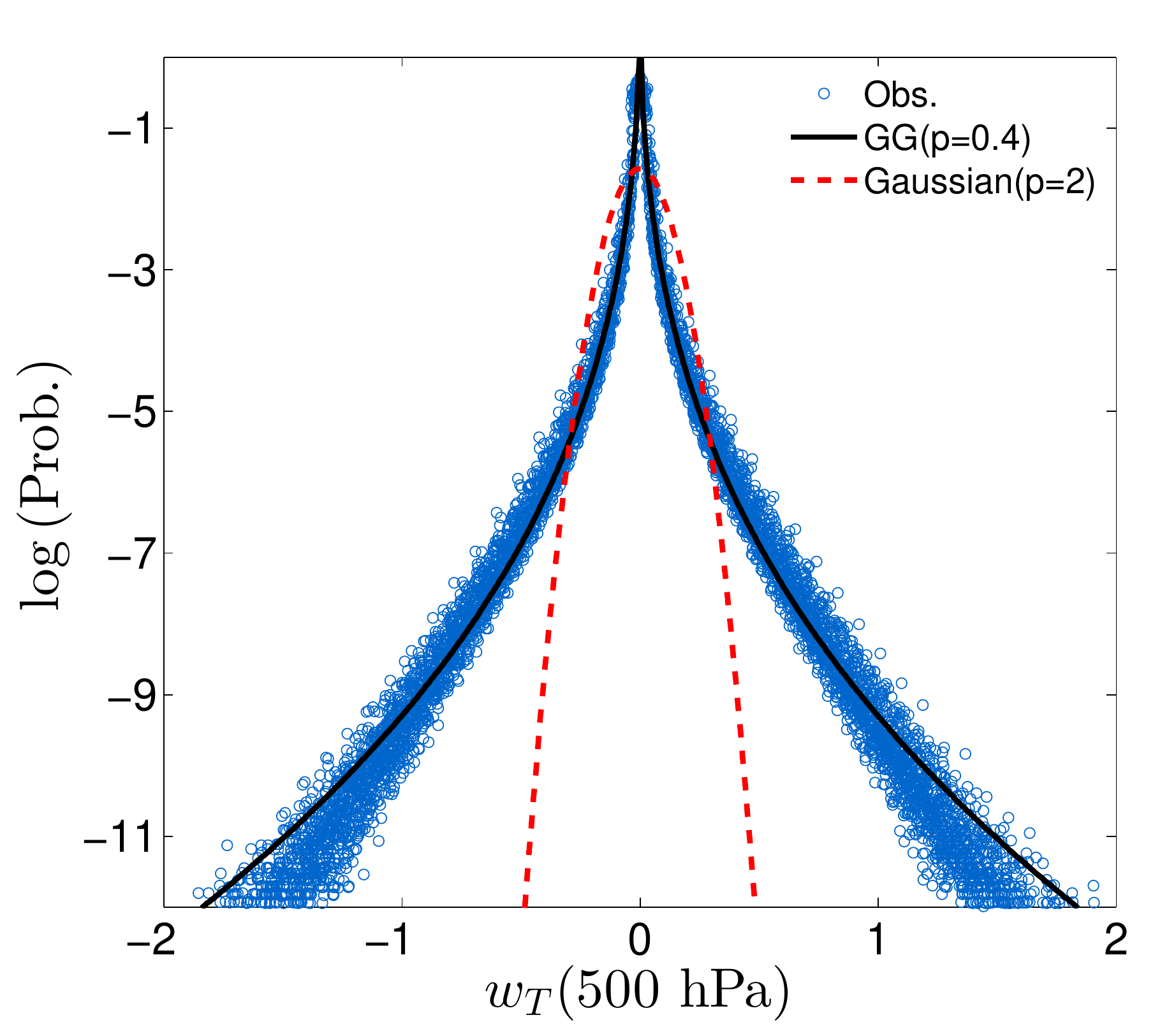} \includegraphics[width=0.35\paperwidth]{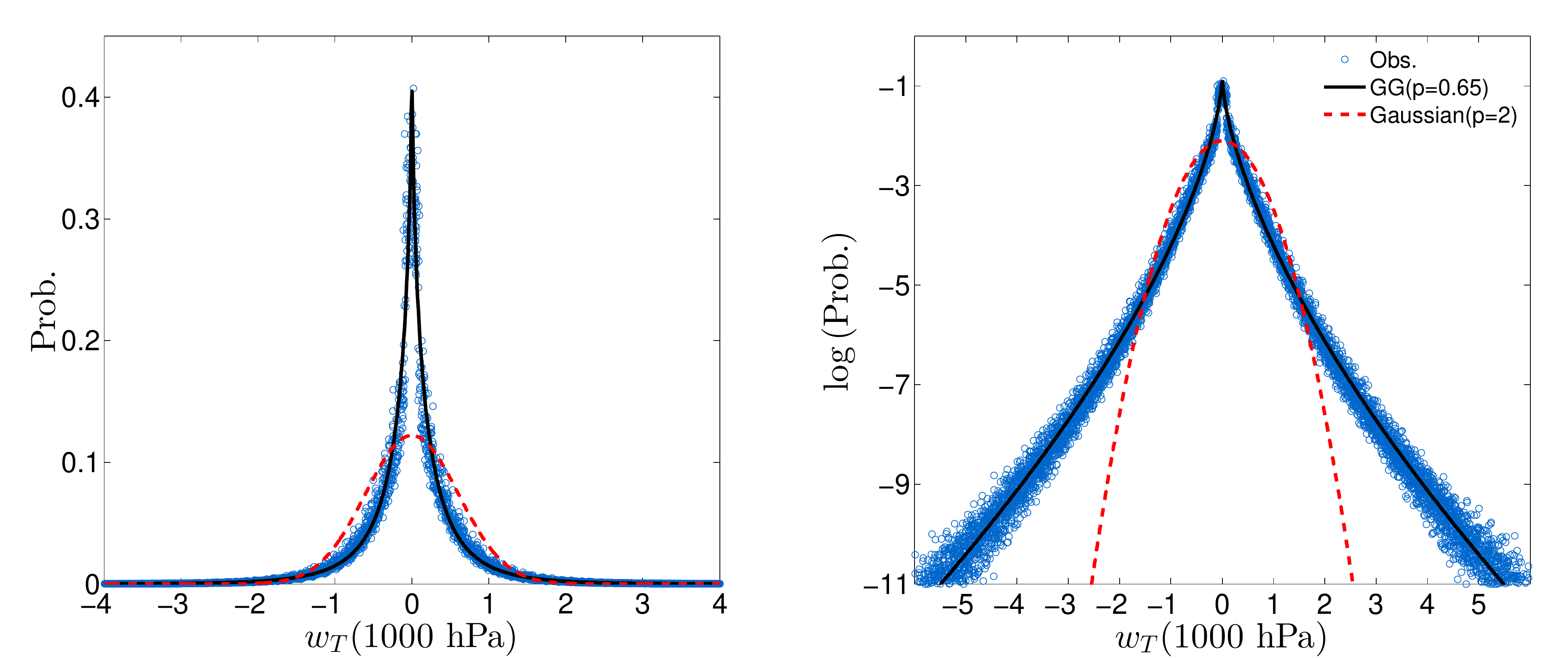}
\par\end{centering}

\protect\caption{Sparsity in the moisture fields. The overlaid empirical probability
masses (circles) of the wavelet coefficients for the fields of the
water vapor mass mixing ratio {[}g/kg dry air{]} at pressure levels
500 hPa (left panel) and 1000 hPa (right panel). Please see caption
of Figure \ref{fig:S2} for relevant explanations.\label{fig:S4}}

\end{figure}

Here, Table \ref{tab:S1} shows
the median and 95\% confidence bound of the computed shape parameters
of the GG distribution, from the moments shown in Figure \ref{fig:3}. Furthermore, supplementary Figure \ref{fig:S1} shows the frequency histogram of the days for which the standard AIRS level II retrievals are used in this study.
Figures \ref{fig:S2}, \ref{fig:S3}, and \ref{fig:S4} provide further
evidence that the fields of geopotential height, temperature and
moisture are sparse in the wavelet domain and their representation
coefficients can be well explained by the Generalized Gaussian (GG)
distribution.


\begin{table}[h]
\noindent \begin{centering}
{\footnotesize{}}%
\begin{tabular}{|c|c|c|c|c|c|c|}
\hline 
 & {GP\_Height} & {SurfSkinTemp} & {SurfAirTemp} & {TAirStd} & {H2OMMRSurf} & {H2OMMRStd}\tabularnewline
\hline 
\hline 
{Median} & {0.63} & {0.50} & {0.59} & {0.76} & {0.61} & {0.40}\tabularnewline
\hline 
{$\text{CI}_{95}$} & {$0.51-0.70$} & {$0.48-0.53$} & {$0.54-0.63$} & {$0.64-0.84$} & {$0.56-0.67$} & {$0.33-0.66$}\tabularnewline
\hline 
\end{tabular}
\par\end{centering}{\footnotesize \par}

\protect\caption{The median and the 95\% confidence interval (${\rm CI}_{95}$) for
the computed shape parameters of the Generalized Gaussian density,
from the moments of the wavelet coefficients shown in Figure \ref{fig:3}. The shape parameters
are obtained by solving equation (\ref{eq:2}) for $p$ using the bisection
method.\label{tab:S1} }
\end{table}

\section{Sparsity Prior and Compressive Sensing \label{sec:S2}}

In this section, we provide more details on how sparsity can be characterized
as a prior knowledge and explain the statistical interpretation of
the CS problem in equation (\ref{eq:5}). To be more precise, let us assume
that a finite dimensional land-atmospheric state vector $\mathbf{x}\in\mathfrak{R}^{m}$
can be represented by its wavelet coefficients $\mathbf{c}\in\mathfrak{R}^{d}$
obtained through a matrix-vector multiplication, that is $\mathbf{c}=\mathbf{W}\mathbf{x}$,
where the rows of $\mathbf{W}\in\mathfrak{R}^{d\times m}$ contain
the wavelet bases or frames of choice for which $\mathbf{W}^{T}\mathbf{W}=\mathbf{I}$,
where $\left(\cdotp\right)^{T}$ denotes transposition. Assuming that
the coefficients are independent and can be well explained by the
GG density, it implies that the prior distribution of the state vector
of interest admits the following multivariate form:

\begin{equation}
\log\, p\left(\mathbf{x}\right)\propto-\lambda\left\Vert \mathbf{W}\mathbf{x}\right\Vert _{p}^{p},\tag{S1}\label{eq:S1}
\end{equation}
where the $\ell_{p}$-norm is $\left\Vert \mathbf{x}\right\Vert _{p}^{p}=\Sigma_{i=i}^{m}\left|x_{i}\right|^{p}$
and $\lambda>0$ collectively encodes the width parameter of the density.
Clearly, from the Bayesian perspective, this a priori knowledge can
be used to obtain an a posteriori estimate of the land-atmospheric
state of interest. Specifically, let us assume that a set of under-sampled
observations $\mathbf{y}\in\mathfrak{R}^{n}$ are related to the true
state $\mathbf{x}\in\mathfrak{R}^{m}$, that is $n<m$, via the following
linear model

\begin{equation}
\mathbf{y=}\mathbf{Hx}+\mathbf{v},\tag{S2}\label{eq:S2}
\end{equation}
where $\mathbf{v}\in\mathfrak{R}^{n}$ is a length-n error vector
of the Gaussian nature with an $n$-by-$n$ covariance matrix $\mathbf{R}\in\mathfrak{R}^{n\times n}$,
and $\mathbf{H}\in\mathfrak{R}^{n\times m}$ denotes the sensing matrix.
The linear system of equations in (\ref{eq:S2}) is under-determined
for which the maximum likelihood estimator does not yield a unique
solution. Clearly, given the observation y, the maximum a posterior
(MAP) estimate of the true state can be derived as follows:

\begin{equation}
\begin{aligned}\mathbf{\hat{\mathbf{x}}}_{\text{MAP}} & =\underset{\mathbf{x}}{\text{argmax}}\; p\left(\mathbf{x}|\mathbf{y}\right)\\
 & =\underset{\mathbf{x}}{\text{argmin}}\;\left\{ -\log\, p\left(\mathbf{y}|\mathbf{x}\right)-\log\, p\left(\mathbf{x}\right)\right\} .
\end{aligned}
\tag{S3}\label{eq:S3}
\end{equation}

Thus, one can easily see that $\hat{\mathbf{x}}_{\text{MAP}}$ can
be obtained by solving the following minimization problem

\begin{equation}
\underset{\mathbf{x}}{\text{minimize}}\;\left\{ \left\Vert \mathbf{y}-\mathbf{H}\mathbf{x}\right\Vert _{\mathbf{R}}^{2}+\lambda\left\Vert \mathbf{W}\mathbf{x}\right\Vert _{p}^{p}\right\} ,\tag{S4}\label{eq:S4}
\end{equation}
or alternatively by solving its counterpart in the wavelet domain 

\begin{equation}
\underset{\mathbf{c}}{\text{minimize}}\;\left\{ \left\Vert \mathbf{y}-\mathbf{H}\mathbf{W}^{{T}}\mathbf{c}\right\Vert _{\mathbf{R}}^{2}+\lambda\left\Vert \mathbf{c}\right\Vert _{p}^{p}\right\} ,\tag{S5}\label{eq:S5}
\end{equation}
where $\left\Vert \mathbf{x}\right\Vert _{\mathbf{R}}^{2}=\mathbf{x}^{{T}}\mathbf{R}^{-1}\mathbf{x}$.
By obtaining the MAP estimate of the wavelet coefficients $\hat{\mathbf{c}}_{{\rm MAP}}$
from problem (\ref{eq:S5}), obviously one can compute the state of
interest as $\hat{\mathbf{x}}_{{\rm MAP}}=\mathbf{W}^{T}\mathbf{\,\hat{c}}_{{\rm MAP}}$.
A proper choice of the prior term, often called regularization, may
allow us to obtain a unique or stable solution for the under-determined
system of equations in (\ref{eq:S2}). Notice that the sparsest solution
of problem (\ref{eq:S5}) can be naturally achieved in case of $p\rightarrow0$;
however, choices of $p<1$ make this problem non-convex. Therefore,
$p=1$ is the smallest value of the shape parameter that promotes
sparsity while maintaining the problem in the realm of convex optimization.
For the CS experiments, we used the Fast Iterative Shrinkage-Thresholding
Algorithm in \citep{BecT09a} to solve the following problem in the
wavelet domain:

\begin{equation}
\underset{\mathbf{c}}{\text{minimize}}\;\left\{ \left\Vert \mathbf{y}-\mathbf{H}\mathbf{W}^{{T}}\mathbf{c}\right\Vert _{\mathbf{R}}^{2}+\lambda\left\Vert \mathbf{c}\right\Vert _{1}\right\} ,\tag{S6}\label{eq:S6}
\end{equation}
and then recovered the state of interest in ambient domain from the estimated coefficients. As explained in the main body of the paper, for further theoretical explanation of random sampling and it implication for the $\ell_1$-norm reconstruction, the reader may refer to the original work by \citet{CandTao05}, while more practical applications in the magnetic resonance imaging can be found in \citep{Lus08}. 

%----------------------------------------------------------------------------------------
%	BIBLIOGRAPHY
%----------------------------------------------------------------------------------------


\begin{thebibliography}{}

\providecommand{\natexlab}[1]{#1}
\expandafter\ifx\csname urlstyle\endcsname\relax
  \providecommand{\doi}[1]{doi:\discretionary{}{}{}#1}\else
  \providecommand{\doi}{doi:\discretionary{}{}{}\begingroup
  \urlstyle{rm}\Url}\fi

\bibitem[{\textit{Beck and Teboulle}(2009)}]{BecT09a}
Beck, A., and M.~Teboulle (2009), {A} {F}ast {I}terative
  {S}hrinkage-{T}hresholding {A}lgorithm for {L}inear {I}nverse {P}roblems,
  \textit{SIAM J. Imaging Sci.}, \textit{2}(1), 183--202,
  \doi{10.1137/080716542}.

\bibitem[{\textit{Bodmann et~al.}(2011)\textit{Bodmann, Casazza, and
  Kutyniok}}]{BodCK11}
Bodmann, B.~G., P.~G. Casazza, and G.~Kutyniok (2011), A quantitative notion of
  redundancy for finite frames, \textit{Appl. Comput. Harmonic Anal.},
  \textit{30}(3), 348 -- 362, \doi{10.1016/j.acha.2010.09.004}.

\bibitem[{\textit{Candes and Romberg}(2006)}]{CanR06}
Candes, E.~J., and J.~Romberg (2006), Quantitative robust uncertainty
  principles and optimally sparse decompositions, \textit{Found. Comput.
  Math.}, \textit{6}(2), 227--254, \doi{10.1007/s10208-004-0162-x}.

\bibitem[{\textit{Candes and Tao}(2005)}]{CandTao05}
Candes, E.~J., and T.~Tao (2005), Decoding by linear programming, \textit{IEEE
  Trans. Inform. Theory.}, \textit{51}(12), 4203--4215,
  \doi{10.1109/TIT.2005.858979}.

\bibitem[{\textit{Candes and Tao}(2006)}]{CanT06}
Candes, E.~J., and T.~Tao (2006), Near-optimal signal recovery from random
  projections: Universal encoding strategies?, \textit{IEEE Trans. Inform.
  Theory.}, \textit{52}(12), 5406--5425, \doi{10.1109/TIT.2006.885507}.

\bibitem[{\textit{Candes et~al.}(2006{\natexlab{a}})\textit{Candes, Romberg,
  and Tao}}]{CanRT06}
Candes, E.~J., J.~Romberg, and T.~Tao (2006{\natexlab{a}}), Robust uncertainty
  principles: exact signal reconstruction from highly incomplete frequency
  information, \textit{IEEE Trans. Inform. Theory.}, \textit{52}(2), 489--509,
  \doi{10.1109/TIT.2005.862083}.

\bibitem[{\textit{Candes et~al.}(2006{\natexlab{b}})\textit{Candes, Romberg,
  and Tao}}]{CanRT06a}
Candes, E.~J., J.~K. Romberg, and T.~Tao (2006{\natexlab{b}}), Stable signal
  recovery from incomplete and inaccurate measurements, \textit{Commun. Pure
  Appl. Math.}, \textit{59}(8), 1207--1223, \doi{10.1002/cpa.20124}.

\bibitem[{\textit{Donoho}(2006)}]{Don06}
Donoho, D. (2006), {C}ompressed sensing, \textit{IEEE Trans. Inform. Theory.},
  \textit{52}(4), 1289--1306, \doi{10.1109/TIT.2006.871582}.

\bibitem[{\textit{Donoho and Elad}(2003)}]{DonE03}
Donoho, D.~L., and M.~Elad (2003), Optimally sparse representation in general
  (nonorthogonal) dictionaries via $\ell_1$ minimization, \textit{Proceedings
  of the National Academy of Sciences}, \textit{100}(5), 2197--2202,
  \doi{10.1073/pnas.0437847100}.

\bibitem[{\textit{Du et~al.}(2012)\textit{Du, Cooper, and
  Fueglistaler}}]{DuCF12}
Du, J., F.~Cooper, and S.~Fueglistaler (2012), Statistical analysis of global
  variations of atmospheric relative humidity as observed by airs, \textit{J.
  Geophys. Res.}, \textit{117}(D12), \doi{10.1029/2012JD017550}.

\bibitem[{\textit{Duarte et~al.}(2008)\textit{Duarte, Davenport, Takhar, Laska,
  Sun, Kelly, and Baraniuk}}]{DuaDB08}
Duarte, M., M.~Davenport, D.~Takhar, J.~Laska, T.~Sun, K.~Kelly, and
  R.~Baraniuk (2008), Single-pixel imaging via compressive sampling,
  \textit{IEEE Signal. Proc. Mag.}, \textit{25}(2), 83--91,
  \doi{10.1109/MSP.2007.914730}.

\bibitem[{\textit{Ebtehaj et~al.}(2014)\textit{Ebtehaj, Zupanski, Lerman, and
  Foufoula-Georgiou}}]{EbtZLF13}
Ebtehaj, A., M.~Zupanski, G.~Lerman, and E.~Foufoula-Georgiou (2014),
  {V}ariational {D}ata {A}ssimilation via {S}parse {R}egularisation,
  \textit{Tellus A.}, \textit{66}, 2178,
  \doi{http://dx.doi.org/10.3402/tellusa.v66.21789}.

\bibitem[{\textit{Ebtehaj and Foufoula-Georgiou}(2013)}]{EbtF12b}
Ebtehaj, A.~M., and E.~Foufoula-Georgiou (2013), {O}n {V}ariational
  {D}ownscaling, {F}usion and {A}ssimilation of {H}ydro-meteorological
  {S}tates: {A} {U}nified {F}ramework via {R}egularization, \textit{Water
  Resour. Res}, \textit{49}(9), 5944--5963, \doi{10.1002/wrcr.20424}.

\bibitem[{\textit{Eldar and Kutyniok}(2012)}]{EldK12}
Eldar, Y., and G.~Kutyniok (2012), \textit{Compressed Sensing: Theory and
  Applications}, Compressed Sensing: Theory and Applications, Cambridge
  University Press, NY.

\bibitem[{\textit{Entekhabi et~al.}(2010)\textit{Entekhabi, Njoku, O'Neill,
  Kellogg, Crow, Edelstein, Entin, Goodman, Jackson, Johnson, Kimball,
  Piepmeier, Koster, Martin, McDonald, Moghaddam, Moran, Reichle, Shi, Spencer,
  Thurman, Tsang, and Van~Zyl}}]{EntAL10}
Entekhabi, D., E.~Njoku, P.~O'Neill, K.~Kellogg, W.~Crow, W.~Edelstein,
  J.~Entin, S.~Goodman, T.~Jackson, J.~Johnson, J.~Kimball, J.~Piepmeier,
  R.~Koster, N.~Martin, K.~McDonald, M.~Moghaddam, S.~Moran, R.~Reichle, J.-C.
  Shi, M.~Spencer, S.~Thurman, L.~Tsang, and J.~Van~Zyl (2010), The soil
  moisture active passive ({SMAP}) mission, \textit{Proceedings of the IEEE},
  \textit{98}(5), 704--716, \doi{10.1109/JPROC.2010.2043918}.

\bibitem[{\textit{Ferguson and Wood}(2010)}]{FerW10}
Ferguson, C.~R., and E.~F. Wood (2010), An evaluation of satellite remote
  sensing data products for land surface hydrology: Atmospheric infrared
  sounder, \textit{J. Hydrometeor.}, \textit{11}(6), 1234--1262,
  \doi{10.1175/2010JHM1217.1}.

\bibitem[{\textit{Foufoula-Georgiou et~al.}(2014)\textit{Foufoula-Georgiou,
  Ebtehaj, Zhang, and Hou}}]{EfgEZa13}
Foufoula-Georgiou, E., A.~Ebtehaj, S.~Zhang, and A.~Hou (2014), {D}ownscaling
  {S}atellite {P}recipitation with {E}mphasis on {E}xtremes: {A} {V}ariational
  $\ell_1$-{N}rom {R}egularization in the {D}erivative {D}omain,
  \textit{Surv. Geophys.}, pp. 1--19, \doi{10.1007/s10712-013-9264-9}.


\bibitem[{\textit{Freitag et~al.}(2012)\textit{Freitag, Nichols, and
  Budd}}]{FreNB12}
Freitag, M.~A., N.~K. Nichols, and C.~J. Budd (2012), {R}esolution of sharp
  fronts in the presence of model error in variational data assimilation,
  \textit{Quart. J. Roy. Meteor. Soc.}, \textit{139}, 749--757,
  \doi{10.1002/qj.2002}.

\bibitem[{\textit{Hou et~al.}(2013)\textit{Hou, Kakar, Neeck, Azarbarzin,
  Kummerow, Kojima, Oki, Nakamura, and Iguchi}}]{Hou2013}
Hou, A.~Y., R.~K. Kakar, S.~Neeck, A.~A. Azarbarzin, C.~D. Kummerow, M.~Kojima,
  R.~Oki, K.~Nakamura, and T.~Iguchi (2013), The global precipitation
  measurement mission, \textit{Bull. Amer. Meteor. Soc.}, \textit{95}(5),
  701--722, \doi{10.1175/BAMS-D-13-00164.1}.

\bibitem[{\textit{Kim et~al.}(2007)\textit{Kim, Koh, Lustig, Boyd, and
  Gorinevsky}}]{KimKLB07}
Kim, S.-J., K.~Koh, M.~Lustig, S.~Boyd, and D.~Gorinevsky (2007), {A}n
  {I}nterior-{P}oint {M}ethod for {L}arge-{S}cale l1-{R}egularized {L}east
  {S}quares, \textit{IEEE J. Sel. Topics Signal Process.}, \textit{1}(4),
  606--617, \doi{10.1109/JSTSP.2007.910971}.

\bibitem[{\textit{Krahmer and Ward}(2014)}]{Karhmer_Ward14}
Krahmer F., and R.~Ward.
\newblock Stable and robust sampling strategies for compressive imaging.
\newblock {\em IEEE Trans. Image Process.}, 23(2):612--622, Feb
  2014.

\bibitem[{\textit{Le~Marshall et~al.}(2006)\textit{Le~Marshall, Jung, Derber,
  Chahine, Treadon, Lord, Goldberg, Wolf, Liu, Joiner, Woollen, Todling, van
  Delst, and Tahara}}]{LeMetal06}
Le~Marshall, J., J.~Jung, J.~Derber, M.~Chahine, R.~Treadon, S.~J. Lord,
  M.~Goldberg, W.~Wolf, H.~C. Liu, J.~Joiner, J.~Woollen, R.~Todling, P.~van
  Delst, and Y.~Tahara (2006), {I}mproving {G}lobal {A}nalysis and
  {F}orecasting with {AIRS}, \textit{Bull. Amer. Meteor. Soc.}, \textit{87}(7),
  891--894, \doi{10.1175/BAMS-87-7-891}.


\bibitem[\textit{Lustig et~al.}(2008)]{Lus08}
Lustig, M., D.~L. Donoho, J.~M. Santos, J.~M. Pauly (2008), Compressed Sensing MRI,
\textit{IEEE Signal Process. Mag.}, \textit{25}, 72--82, \doi{10.1109/MSP.2007.914728}.

\bibitem[{\textit{Mallat}(1989)}]{Mal89}
Mallat, S. (1989), {A} theory for multiresolution signal decomposition: the
  wavelet representation, \textit{IEEE Trans. Pattern Anal. Mach. Intell.},
  \textit{11}(7), 674--693, \doi{10.1109/34.192463}.

\bibitem[{\textit{Mallat}(2009)}]{Mal09}
Mallat, S. (2009), \textit{{A} wavelet tour of signal processing: the sparse
  way}, 3rd ed., 805 pp., Elsevier /Academic Press, Burlington, MA.

\bibitem[{\textit{Nadarajah}(2005)}]{Nad05}
Nadarajah, S. (2005), {A} generalized normal distribution, \textit{J. Appl.
  Stat.}, \textit{32}(7), 685--694.

\bibitem[{\textit{Nason and Silverman}(1995)}]{NasS95}
Nason, G.~P., and B.~W. Silverman (1995), {T}he {S}tationary {W}avelet
  {T}ransform and some {S}tatistical {A}pplications, \textit{Lecture Notes in
  Statist.}, \textit{103}, 281--299.

\bibitem[{\textit{Olsen et~al.}(2013)}]{Ols13}
Olsen, T.~E., E. Fetzer, G. Hulley, E. Manning, J. Blaisdell, L. Iredell, J. Susskind, J. Warner, Z. Wei, W. Blackwell, E. Maddy (2013), AIRS/AMSU/HSB Version 6 Level 2 Product User Guide, pp. 134, available at \url{http://disc.sci.gsfc.nasa.gov/AIRS/documentation/v6_docs}.

\bibitem[{\textit{Pagano et~al.}(2003)\textit{Pagano, Aumann, Hagan, and
  Overoye}}]{PagAHO03}
Pagano, T., H.~Aumann, D.~Hagan, and K.~Overoye (2003), {P}relaunch and
  in-flight radiometric calibration of the {A}tmospheric {I}nfrared {S}ounder
  ({AIRS}), \textit{IEEE Trans. Geosci. Remote.},
  \textit{41}(2), 265--273, \doi{10.1109/TGRS.2002.808324}.

\bibitem[{\textit{Rauhut and Ward}(2012)}]{RauhutW12}
Rauhut H., and R.~Ward.
\newblock Sparse legendre expansions via l1-minimization.
\newblock {\em Journal of Approximation Theory}, 164(5):517--533, 2012.

\bibitem[{\textit{Reale et~al.}(2008)\textit{Reale, Susskind, Rosenberg, Brin,
  Liu, Riishojgaard, Terry, and Jusem}}]{Realeetal08}
Reale, O., J.~Susskind, R.~Rosenberg, E.~Brin, E.~Liu, L.~P. Riishojgaard,
  J.~Terry, and J.~C. Jusem (2008), {I}mproving forecast skill by assimilation
  of quality-controlled {AIRS} temperature retrievals under partially cloudy
  conditions, \textit{Geophys. Res. Lett.}, \textit{35}(8), L08,809.

\bibitem[{\textit{Reale et~al.}(2009)\textit{Reale, Lau, Susskind, Brin, Liu,
  Riishojgaard, Fuentes, and Rosenberg}}]{ReaEtal09}
Reale, O., W.~K. Lau, J.~Susskind, E.~Brin, E.~Liu, L.~P. Riishojgaard,
  M.~Fuentes, and R.~Rosenberg (2009), {AIRS} impact on the analysis and
  forecast track of tropical cyclone {N}argis in a global data assimilation and
  forecasting system, \textit{Geophys. Res. Lett.}, \textit{36}(6),
  \doi{10.1029/2008GL037122}.

\bibitem[{\textit{Reale et~al.}(2012)\textit{Reale, Lau, Susskind, and
  Rosenberg}}]{ReaLSR12}
Reale, O., K.~Lau, J.~Susskind, and R.~Rosenberg (2012), {AIRS} impact on
  analysis and forecast of an extreme rainfall event ({I}ndus {R}iver {V}alley,
  {P}akistan, 2010) with a global data assimilation and forecast system,
  \textit{J. Geophys. Res.}, \textit{117}, D08,103,
  \doi{doi:10.1029/2011JD017093}.

\bibitem[{\textit{Susskind et~al.}(2003)\textit{Susskind, Barnet, and
  Blaisdell}}]{SusBB03}
Susskind, J., C.~Barnet, and J.~Blaisdell (2003), {R}etrieval of atmospheric
  and surface parameters from {AIRS}/{AMSU}/{HSB} data in the presence of
  clouds, \textit{IEEE Trans. Geosci. Remote.}, \textit{41}(2), 390--409,
  \doi{10.1109/TGRS.2002.808236}.

\bibitem[{\textit{Susskind et~al.}(2011)\textit{Susskind, Blaisdell, Iredell,
  and Keita}}]{SusBIK11}
Susskind, J., J.~Blaisdell, L.~Iredell, and F.~Keita (2011), {I}mproved
  {T}emperature {S}ounding and {Q}uality {C}ontrol {M}ethodology {U}sing
  {AIRS}/{AMSU} {D}ata: {T}he {AIRS} {S}cience {T}eam {V}ersion 5 {R}etrieval
  {A}lgorithm, \textit{IEEE Trans. Geosci. Remote.}, \textit{49}(3), 883--907,
  \doi{10.1109/TGRS.2010.2070508}.

\bibitem[{\textit{Tian et~al.}(2006)\textit{Tian, Waliser, Fetzer, Lambrigtsen,
  Yung, and Wang}}]{Tia06}
Tian, B., D.~E. Waliser, E.~J. Fetzer, B.~H. Lambrigtsen, Y.~L. Yung, and
  B.~Wang (2006), {V}ertical {M}oist {T}hermodynamic {S}tructure and
  {S}patial--{T}emporal {E}volution of the {MJO} in {AIRS} {O}bservations,
  \textit{J. Atmos. Sci.}, \textit{63}(10), 2462--2485,
  \doi{10.1175/JAS3782.1}.

\bibitem[{\textit{Wu et~al.}(2012)\textit{Wu, Su, Fovell, Wang, Shen, Kahn,
  Hristova-Veleva, Lambrigtsen, Fetzer, and Jiang}}]{Wuetal12}
Wu, L., H.~Su, R.~G. Fovell, B.~Wang, J.~T. Shen, B.~H. Kahn, S.~M.
  Hristova-Veleva, B.~H. Lambrigtsen, E.~J. Fetzer, and J.~H. Jiang (2012),
  Relationship of environmental relative humidity with north atlantic tropical
  cyclone intensity and intensification rate, \textit{Geophys. Res. Lett.},
  \textit{39}(20), \doi{10.1029/2012GL053546}.

\end{thebibliography}
\end{document}